\documentclass[twocolumn,pra,superscriptaddress]{revtex4-2}
\usepackage{graphicx,subfigure}          
\usepackage{amsfonts,amsmath,amssymb, amsthm}    
\usepackage{dsfont}                      
\usepackage[dvipsnames]{xcolor}          
\usepackage[
colorlinks=true,
linkcolor=blue,
citecolor=blue,
urlcolor=blue]{hyperref}                
\usepackage{comment}                    
\usepackage{physics}                    
\usepackage[normalem]{ulem}
\usepackage{bm}
\usepackage{mdframed}
\usepackage{enumitem}
\usepackage{soul}

\newcommand{\chg}[1]{\textcolor{purple}{#1}} 

\newtheorem{definition}{Definition}

\begin{document}


\title{Local finite rate semi device independent randomness expansion from quantum contextuality}

\author{Dhananjoy Ghosh}
\email{d25062@students.iitmandi.ac.in}
\affiliation{Center for Quantum Science and Technology, Indian Institute of Technology, Mandi, Himachal Pradesh 175075, India}

\author{Nandana J}
\email{di2408@students.iitmandi.ac.in}
\affiliation{Center for Quantum Science and Technology, Indian Institute of Technology, Mandi, Himachal Pradesh 175075, India}
\affiliation{School of Physical Sciences, Indian Institute of Technology, Mandi, Himachal Pradesh 175075, India}

\author{Jaskaran Singh}
\email{jaskaran@iitmandi.ac.in}
\affiliation{Center for Quantum Science and Technology, Indian Institute of Technology, Mandi, Himachal Pradesh 175075, India}


\begin{abstract}
Finite rate randomness expansion (RE) which is secure against quantum adversaries is a crucial information processing task which is well studied in the case of two non-communicating parties sharing Bell non-local correlations in a device independent manner. In this paper we show that contextual correlations with only a single localized party are sufficient for RE in a semi-device independent (SDI) manner. We propose a template SDI-RE protocol which is applicable for arbitrary $N$-cycle non-contextuality (NC) inequalities and derive its RE rate under finite statistics. As a special case, we also highlight the advantages of the situation when RE is constrained by Clauser-Horne-Shimony-Holt (CHSH) inequality but with characterized measurements implemented by only a single party. Our template protocol also incorporates non-ideal statistics which may not satisfy perfect commutativity and perfect repeatability, both of which are necessary in a loophole-free test of NC inequalities. As such, several recent experimental implementations of significantly loophole-free tests of NC inequalities can be modified for secure RE following our approach.
\end{abstract}


\maketitle


\section{Introduction}
\label{sec:intro}

Randomness expansion (RE) is a crucial task in several cryptographic applications, where a shorter random bit string is used to generate a longer bit string which is uniform and secure against arbitrary adversaries. Quantum theory has proven to be an excellent candidate for RE, even allowing security against computationally unbounded adversaries with access to quantum resources, called as quantum adversaries.

Recently, significant focus has been on device independent RE (DI-RE) protocols~\cite{CK11, FGS13, LZL18, BRC20, LZL21, SZB21, LLR21}, which have been shown to be secure against quantum adversaries and can be universally composed~\cite{C01, PR22} with other secure cryptographic protocols even for finite experimental statistics by employing recently developed techniques like entropy accumulation theorem~\cite{DF19, DFR20, MFS22} (EAT) and probability estimation~\cite{ZKB18, KZF18, XFK20}. This has also led to their experimental realizations~\cite{LLR21, LZL21, LYL18, SZB21}.


One of the key advantages of a DI-RE protocol is that it requires minimal characterization of the devices under use. Instead, it leverages input-output statistics, typically in the form of a Bell inequality~\cite{B64, CHSH69}, to certify security. Real-world devices are susceptible to implementation errors and imperfections that theoretical models may fail to capture, potentially leading to an overestimation of security, which necessitates their proper characterization during their operation. As a result this approach is particularly valuable.

However, DI-RE protocols necessitate at least two non-communicating parties which should violate a Bell inequality in a loophole-free fashion~\cite{P70, E93}. This is a very stringent requirement which is experimentally hard to achieve~\cite{ VPB10, RKM01, GMR13, CMK13,SMC15, GVW15, HBD15, RBG17, XSS23, SA24}, resulting in lower randomness generation rates.

On the other hand, several RE protocols have also been developed in which either the source or the measurement apparatus is fully characterized while the other is not~\cite{CZM15,NGZ16, BKB17,XSW16, HWCGP17, ATVV21, WBGB22,SSC17, BME17, TZASVV21,CZY16, AMV18, DWH20}. These protocols, called semi-device independent (SDI), can theoretically offer higher RE rates as compared to DI-RE protocols, but at the cost of slightly reduced security claims. This trade-off is particularly advantageous in scenarios that demand high rates of randomness generation. On the downside, a comprehensive finite-rate security analysis for SDI-RE protocols is available for only a limited number of cases~\cite{CZY16, AMV18, DWH20}. 

With the growing demand for quantum-secure technologies, it is natural to develop RE protocols that can enjoy the benefits of both DI-RE and SDI-RE schemes by identifying the best qualities of both. A `good' RE protocol should (i) rely only on a single localized party, (ii) be experimentally feasible and (iii) remain secure against quantum adversaries even under the effects of finite experimental rounds. Such protocols should also be universally composable with other cryptographic schemes and should also offer the advantage of certifying security based on some input-output statistics which does not waste too many resources on their characterization.

In this paper we show that quantum contextuality~\cite{CSW14, budroni2022kochen, AQB13, SBA23} is one such resource that has the potential to offer the aforementioned qualities in an RE protocol. 
Much like Bell non-local correlations, contextual correlations can be certified by a violation of a statistical inequality that captures the non-contextual behavior of the underlying theory, termed as a non-contextuality (NC) inequality~\cite{CSW14, SBA23, SBA23_2}. 
Contextual correlations are already instrumental in several information processing tasks like quantum computation~\cite{HWV14}, quantum key distribution~\cite{SBA17}, local self-testing~\cite{BRV19, BRX22, HXA23}, dimension certification~\cite{ray2021graph} and more recently in RE using self-testing~\cite{SFB24}.

We propose a template SDI-RE protocol using a single localized party that does not characterize the source, while the measurements are only partially characterized. Specifically, the security is certified by the violation of a state-dependent NC inequality~\cite{AQB13}, with some key assumptions (such as memorylessness, repeatability, measurement compatibility) aligning with their loophole-free testing~\cite{Kirchmair:2009NAT, WZL22, PJC15, CLB15}. The security proof of our proposed scheme follows the standard definition of universal composability which also allows our protocol to be implemented in conjunction with any other (secure) cryptographic task without compromising the overall security. 

We generalize our scheme to violation of arbitrary, but state-dependent contextuality scenarios. For illustration we explicitly provide numerical results corresponding to the Klyachko-Can-Binicio{\u g}lu-Shumovsky (KCBS) inequality~\cite{KCB08} and the Clauser-Horne-Shimony-Holt (CHSH) inequality~\cite{CHSH69}. For our purposes we re-interpret the latter scenario as a contextuality scenario in which the state and measurements are implemented by a single party and satisfy the no-disturbance principle (instead of no-signaling). This re-interpretation also highlights the usefulness of the CHSH scenario for RE when the measurements are assumed to be partially characterized as opposed to DI-RE scenarios in which they are fully uncharacterized.


Our current work also significantly improves upon an earlier scheme~\cite{SFB24} which relied on close to maximal violation of NC inequalities for a local self-test and could only function for an asymptotically infinite number of experimental rounds. Instead we provide a finite rate security analysis for arbitrary violations of state-dependent NC inequalities.

The paper is organized as follows. In Sec.~\ref{sec:context} we provide the necessary background details on quantum contextuality. In Sec.~\ref{sec:protocl} we lay out the template protocol for RE based on two different classes of state-dependent NC inequalities. In Sec.~\ref{sec:results} we discuss numerical simulations of our template protocols based on the KCBS and CHSH inequality. In Sec.~\ref{sec:methods} we provide a detailed methodology on deriving the secure RE rates. In Sec.~\ref{sec:loopholefree} we discuss how experimental deviations in characterization of measurement devices will affect the secure RE rate analysis. In Sec.~\ref{sec:conc} we provide concluding remarks.


\section{Quantum contextuality}
\label{sec:context}

In this section we provide a brief background on the graph theoretic framework of quantum contextuality~\cite{CSW14} with a focus on $N$-cycle NC inequalities.


We consider a finite set of measurements represented in quantum theory by dichotomic projectors $\{\Pi_x\}_{x=0}^{N - 1}$ having outcomes denoted by $a\in \lbrace 0,1 \rbrace$. Let $\mathcal{G} = (\mathcal{V}, \mathcal{E})$ be the graph of compatibility of $\{\Pi_x\}_{x=0}^{N - 1}$, where each of the elements of the set of vertices $\mathcal{V} = \{v_x\}_{x=0}^{N-1}$ represents one element of $\{\Pi_x\}_{x=0}^{N-1}$, and there is an edge $(v_x, v_y) \in \mathcal{E} \subseteq \mathcal{V} \times \mathcal{V}$ if and only if $\Pi_x$ and $\Pi_y$ are jointly measurable or compatible. 

To illustrate our results, we first focus on a particular set of graphs which have an odd number of vertices connected by edges cyclically~\cite{AQB13}. For these set of graphs $\mathcal{G}$ with $N \ge 5$ and $N$ odd, the following NC inequality is satisfied by any non-contextual hidden-variable (NCHV) theory \cite{AQB13}:
\begin{equation}
    \mathcal{\beta} := \frac{1}{N}\sum_{x = 0}^{N-1} p(1,0|x,x+1) \leq \beta_{\text{NCHV}},
    \label{eq:nc_inequality}
\end{equation}
where $p(1,0|x,x+1)$ is the probability of obtaining outcomes $1$ and $0$ for $\Pi_x$ and $\Pi_{x+1}$ respectively, the sum in $x+1$ is taken modulo $N$
and $\beta_{\text{NCHV}}$ is the maximum of $\beta$ for NCHV theories. In this case, $\beta_{\text{NCHV}} = \frac{\alpha(\mathcal{G})}{N}$, where $\alpha(\mathcal{G})$ is the independence number of the graph $\mathcal{G}$~\cite{CSW14}, while
in quantum theory the maximum of $\beta$ is $\beta_{\text{QT}} = \frac{\vartheta(\mathcal{G})}{N}$, where $\vartheta(\mathcal{G})$ is the Lov{\'a}sz number of the graph $\mathcal{G}$~\cite{CSW14}. More specifically, for the odd-cycle NC inequalities with $N\geq 5$ odd, these bounds take the form
\begin{equation}
    \beta_{\text{NCHV}} = \frac{N-1}{2N},
\end{equation}
and
\begin{equation}
    \beta_{\text{QT}} = \frac{\cos(\pi/N)}{1+\cos(\pi/N)}.
\end{equation}

A particular quantum strategy to realize $\beta_{\text{QT}}$ in an experiment comprises of rank-$1$ projective measurements acting on a pure qutrit state. As an example consider the qutrit state
\begin{equation}
    \ket{\psi} = \left(1, 0, 0\right)^T,
\label{eq:cyclic_strategy}
\end{equation}
where $T$ indicates transposition, and the projective measurements 
\begin{equation}
    \Pi_x = \ket{v_x} \bra{v_x},\;\text{with }x \in \lbrace 0, \ldots, N-1\rbrace,
    \label{eq:projectors}
\end{equation}
where
\begin{equation}
    \ket{v_x} = \left(\cos\theta, \sin\theta \sin\phi_x, \sin\theta \cos\phi_x\right)^T,
\end{equation}
with $\cos^2\theta = \frac{\cos \left(\pi/N\right)}{1 + \cos \left(\pi/N\right)}$ and $\phi_x = \frac{x \pi(N - 1)}{N}$. Note that the projective measurements are chosen to be cyclically orthogonal such that $\Pi_x$ and $\Pi_{x + 1}$ are jointly measurable. Furthermore, we term the system to be a localized system in the sense that all measurements are performed on a single indivisible quantum state.

For $N \ge 4$, with $N$ even, the NC inequalities are different from their odd counterparts. As shown in Ref.~\cite{AQB13} these NC inequalities can be written as
\begin{equation}
    \begin{aligned}
        \mathcal{\beta}' := &\frac{1}{N}\Bigg[ \sum_{x = 0}^{N-2} p(0,0|x,x+1) + p(1,1|x,x+1) \\
        &\quad\quad + p(0,1|N-1,0) + p(1,0|N-1,0)\Bigg] \leq \beta'_{\text{NCHV}},
    \end{aligned}
\label{evenineq}
\end{equation}
where the NCHV bound is
\begin{equation}
\beta'_{\text{NCHV}} = \alpha(M_{2N}) = \frac{N-1}{N},
\end{equation}
where $M_{2N}$ is the M\"obius ladder of order $2N$ \cite{AQB13}, which is the graph of exclusivity of the $2N$ events in \eqref{evenineq}. It should be noted that the exclusivity graph of these scenarios is not the same as the compatibility graph, which is cyclic.
The maximum quantum value is
\begin{equation}
\beta'_{\text{QT}} = \vartheta(M_{2N}) = \frac{1}{2} \left[1 + \cos \left(\pi/N\right)\right].
    \label{eq:cyclic_lovasz}
\end{equation}

$\beta'_{\text{QT}}$ can be achieved \cite{P70} with the two-qubit state
\begin{equation}
    \ket{\psi} = \frac{1}{\sqrt{2}}\left(1,0,0,1\right)^T,
\end{equation}
and the measurements 
\begin{equation}
    \Pi_x = \ket{v_x}\!\bra{v_x} = \frac{1}{2}\left[\mathds{1} + \bm{S}(x)\cdot\bm{\sigma}\right],
    \label{eq:meas_bell}
\end{equation}
which acts on first qubit if $x \in \lbrace 0, 2, \ldots, N-2 \rbrace$ and on the second qubit if $x \in \lbrace 1, 3, \ldots,  N - 1 \rbrace$, $\bm{S}(x) = \left(\cos \left(x \pi/N\right), \sin \left(x \pi/N\right), 0\right)^T$ and $\bm{\sigma} = \left(\sigma_x, \sigma_y, \sigma_z\right)$.

It is worth noting here that while even $N$-cycle NC inequalities necessitate a two-qubit quantum strategy, we consider that both the qubits form a single system. This is distinct from Bell non-local scenarios in which the two qubits are spatially separated and it is required that the corresponding quantum strategies should ensure no-communication. Instead, here, we consider that the quantum strategy is implemented on a localized system and satisfies the no-disturbance principle (see e.g. Ref.~\cite{RSK12} for a note on the no-disturbance principle). Such contextuality scenarios can (and have been) experimentally implemented in a significantly loophole-free fashion~\cite{Kirchmair:2009NAT, WZL22, PJC15, CLB15}.

\section{Local randomness expansion protocol}
\label{sec:protocl}

In this section we first detail a template protocol for RE using $N$-cycle NC inequalities~\cite{AQB13}. Thereafter we proceed to illustrate RE using $N = 5$ and $N = 4$ NC inequalities. 

\begin{figure}
\begin{mdframed}
{\bf Parameters and notation:}
\begin{itemize}[leftmargin = 0.3cm]
    \item[] $n \in \mathbb{N}$ - Total number of rounds.  

    \item[] $N\geq 4\in \mathbb{N}$ - Total number of measurements, which is at least $4$.

    \item[] $\Pi_x$, $x\in \lbrace 0, \ldots, N-1\rbrace$ - Two-outcome projective measurements with outcomes labeled by $a \in \lbrace 0, 1\rbrace$ corresponding to $\mathds{1} - \Pi_x$ and $\Pi_x$ respectively.

    \item[] $\rho$ - Initial quantum state on which the measurements are performed.

    \item[] $\beta_Q$ - Expected value of $N$-cycle NC inequality.

    \item[] $q \in \left[0,1\right]$ and $T \in \lbrace 0,1\rbrace$.

    \item[] $\delta \in (0,1)$ - Confidence interval.

    \item[] $C_j \in \lbrace \perp, 0, 1\rbrace$.

    \item[] $F_C(\mathcal{G})$ - Emperical frequency distribution corresponding to the NC inequality over the graph $\mathcal{G}$.

    \item[] Ext - Quantum-proof randomness extractor.

    \item[] $\textbf{s}$ - Random seed for randomness extraction.  
\end{itemize}
\hrule
\vspace{0.5mm}
\hrule
\vspace{0.2cm}
{\bf Protocol for QRE - Odd $N$-cycle}
\vspace{0.2cm}
\hrule
\vspace{0.2cm}
\textbf{Procedure}
\begin{enumerate}[leftmargin = 0.3cm]
    \item Alice chooses a quantum strategy $\left(\rho, \lbrace \Pi_i\rbrace_{i = 1}^{N}\right)$ corresponding to the quantum realization of the odd $N$-cycle NC inequality and sets $j = 1$.
    \item \textbf{While} $j \leq n$:
    \begin{itemize}
        \item[] For $q \in \left[0, 1\right]$, choose $T_j = 0$ with probability $1 - q$ and $T_j = 1$ otherwise. 

        \item[] \textbf{If} $T_j = 0$: (Generation round) 
        
        \begin{itemize}
            \item[] Perform the measurement $\Pi_{x^*}$ on state $\rho$ to obtain outcome $a$.
            
            \item[] Record $a$ as $k_j$.
            
            \item[] Set $C_j = \perp$.
            
        \end{itemize}

        \item[] \textbf{Else}: (Spot-check round)
        \begin{itemize}
            \item[] Randomly choose $x \in \lbrace 0, \ldots, N-1\rbrace$ and $y = x + 1$ (modulo $N$).
        
            \item[] Perform the measurement $\Pi_x$ on state $\rho$ to obtain outcome $a$. Perform the measurement $\Pi_{y}$ on the post-measurement state to obtain outcome $a'$. Record $a, a', x$.

            \item[]\textbf{If} $a=1$ and $a'=0$:
                \begin{itemize}
                    \item[] Set $C_j = 1$. 
                \end{itemize}
            \item[]\textbf{Else}:
                \begin{itemize}
                    \item[] Set $C_j = 0$. 
                \end{itemize}
                
        \end{itemize} 
        
        \item[] Set $j = j+1$.
    \end{itemize}

    \item Using the statistics from all spot-check rounds evaluate $F_C(\mathcal{G})$.

    \item[] \textbf{If} $q \left(\beta_Q - \delta\right) \leq F_C(\mathcal{G}) \leq q \left(\beta_Q + \delta\right)$:
    \begin{itemize}
        \item[] Obtain the random bit string $\bm{k}$ as a concatenation of all bit values $k_j$ from the generation rounds.

        \item[] Obtain the final random bit string $\bm{z} = {\rm Ext}(\bm{k}, \textbf{s})$.
    \end{itemize}
    
    \item[] \textbf{Else}:
    \begin{itemize}
        \item[] Abort the protocol.
    \end{itemize}
    
\end{enumerate}
\end{mdframed}
\caption{The protocol for RE using contextual correlations in odd $N$-cycle scenarios.}
\label{fig:protocol_odd}
\end{figure}

\begin{figure}
\begin{mdframed}
{\bf Protocol for QRE - Even $N$-cycle}
\vspace{0.2cm}
\hrule
\vspace{0.25cm}

\textbf{Procedure}
\begin{enumerate}[leftmargin = 0.3cm]
    \item Alice chooses a quantum strategy $\left(\rho, \lbrace \Pi_i\rbrace_{i = 1}^{N}\right)$ corresponding to the quantum realization of the even $N$-cycle NC inequality and sets $j = 1$.
    \item \textbf{While} $j \leq n$:
    \begin{itemize}
        \item[] For $q \in \left[0, 1\right]$, choose $T_j = 0$ with probability $1 - q$ and $T_j = 1$ otherwise. 

        \item[] \textbf{If} $T_j = 0$: (Generation round) 
        
        \begin{itemize}
            \item[] Perform the measurement $\Pi_{x^*}$ on state $\rho$ to obtain outcome $a$.
            
            \item[] Record $a$ as $k_j$.
            
            \item[] Set $C_j = \perp$.
            
        \end{itemize}

        \item[] \textbf{Else:} (Spot-check round)
        \begin{itemize}
            \item[] Randomly choose $x \in \lbrace 0, \ldots, N-1\rbrace$ and $y = x + 1$ (modulo $N$).
            Perform the measurement $\Pi_x$ on state $\rho$ to obtain outcome $a$. Perform the measurement $\Pi_{y}$ on the post-measurement state to obtain outcome $a'$. Record $a, a', x$.

\begin{description}
    \item[\textbf{If} $x \in \{0,\ldots,N-2\}$] 
    \noindent
    \begin{itemize}
        \item[] \textbf{If} $a = a'$: 

         set $C_j = 1$.
        \item[]\textbf{Else}:
        
        set $C_j = 0$.
    \end{itemize}
    \item[\textbf{Else}] \noindent
    \begin{itemize}
        \item[]\textbf{If} $a \neq a'$: 

         set $C_j = 1$.
        \item[] \textbf{Else}:
        
        set $C_j = 0$.
    \end{itemize}
\end{description}

        \end{itemize} 
        
        \item[] Set $j = j+1$.
    \end{itemize}

    \item Using the statistics from all spot-check rounds evaluate $F_C(\mathcal{G})$.

    \item[] \textbf{If} $q \left(\beta_Q - \delta\right) \leq F_C(\mathcal{G}) \leq q \left(\beta_Q + \delta\right)$:
    \begin{itemize}
        \item[] Obtain the random bit string $\bm{k}$ as a concatenation of all bit values $k_j$ from the generation rounds.

        \item[] Obtain the final random bit string $\bm{z} = {\rm Ext}(\bm{k}, \textbf{s})$.
    \end{itemize}
    
    \item[] \textbf{Else}:
    \begin{itemize}
        \item[] Abort the protocol.
    \end{itemize}
    
\end{enumerate}
\end{mdframed}
\caption{The protocol for RE using contextual correlations in even $N$-cycle scenarios.}
\label{fig:protocol_even}
\end{figure}

We consider a single party Alice who wishes to generate more randomness than consumed, a task termed as randomness expansion (RE). She has access to a device with inputs $x \in \lbrace 0, \ldots, N-1 \rbrace$ ($N \geq 4$) such that each of the inputs can yield two outcomes labeled by $a \in \lbrace 0, 1 \rbrace ~\forall x$. 

A particular implementation of Alice's device can be provided by considering a quantum strategy in which Alice possesses a quantum state $\rho_A$, upon which the device implements the observable $A_x = M_{0|x} - M_{1|x}$, where $\lbrace M_{a|x} \rbrace_{a}$ are the outcomes corresponding to the measurement for setting $x$.

In our protocol we consider that the measurement devices of Alice are partially characterized while the preparation source is untrusted. This constitutes an SDI scenario. The partial characterization of Alice's measurements entails checking how close they are to being repeatable and compatible (following the graph $\mathcal{G}$). The condition of repeatability of measurements ensures that Alice observes the same outcome when the same measurement is implemented twice sequentially, while the condition of compatibility ensures that measurements, according to the graph $\mathcal{G}$, are orthogonal. We provide more details on this in Sec.~\ref{sec:loopholefree} 


We also assume that all the devices of Alice are memoryless such that the state prepared in each experimental round is independent of all previous rounds, and measurement outcomes do not depend on prior settings or outcomes. As a result all experimental rounds are assumed to be independent of each other but may not be identical. This is a necessary assumption for ensuring a loophole-free violation of NC inequalities. 

Furthermore, we also assume that Alice has access to a private string of bits which will be consumed to run the RE protocol. 

We also consider a computationally unbounded adversary, Eve who may have access to Alice's preparation device and can also potentially establish non-trivial correlations with her outcomes while remaining undetected. Specifically, we assume that Eve holds a purification $\ket{\psi}_{AE}$ of $\rho_A$. Since Alice will generate the raw bit string from the outcomes of a single input setting, say $X = x$, the resultant post-measurement joint correlations of Alice and Eve can be described by a classical-quantum (cq) state $\rho_{AE|X = x} = \sum_{a} \ket{a} \bra{a}_A \otimes \rho_{E}^{a|x}$, where $\rho_{E}^{a|x}$ is the reduced state of Eve when Alice obtains the outcome $a$ conditioned on the setting $x$, such that the observed correlations with Alice satisfy $p(a|x) = \text{tr}\left[\left(M_{a|x} \otimes \mathds{1}_E\right)\ket{\psi}\bra{\psi}_{AE}\right]$ $\forall a, x$. 

For our purposes, we focus on the class of spot-checking protocols for RE. Qualitatively, this class of protocols can be understood as having two major steps (see e.g. Ref.~\cite{BRC20}), but for the purpose of an SDI protocol in which the measurement devices are partially characterized we include an additional initial step which is implemented prior to executing the actual protocol. These steps can be briefly described as follows. 
\begin{enumerate}
    \item \textbf{Measurement device characterization:} In this step Alice characterizes her measurement device to ensure that the measurements are repeatable and follow the necessary compatibility relations according to the graph $\mathcal{G}$. Since noise and imperfections in the devices are inevitable, small deviation from perfect repeatability and compatibility would be observed.
    Alice quantifies this deviation from ideal statistics which then determines the confidence interval of the NC inequality violation that will be used in the execution of the actual protocol. As such it will also affect the probability with which the protocol aborts. We provide more details of this step, the necessary characterizations, quantification of the deviation from ideal statistics and its influence on the chosen confidence interval to observe a violation in Sec.~\ref{sec:loopholefree}.
    
    \item \textbf{Spot-check and raw randomness accumulation:} In this step Alice uses a short initial random bit string to decide whether an experimental round will be a spot-check or generation round. In the spot-check rounds, Alice randomly selects a measurement setting $x$ and the setting $x +1$ (modulo $N$), which are implemented sequentially to obtain the outcomes $a$, $a'$. The statistics collected from all spot-check rounds is used to check for the violation of the NC inequality. On the other hand, in each generation round, Alice uses a pre-defined setting $x^*$ to obtain the outcome $a$, which is concatenated to obtain the random bit string $\bm{k}$ only if the input-output statistics in the spot-check rounds violate an NC inequality. In case a violation is not observed, the protocol is aborted and Alice would need to begin anew.

    \item \textbf{Extraction:} In this step (only when the statistics from the spot-check rounds indicate a violation of the NC inequality), Alice applies a strong quantum-proof randomness extractor to the string $\bm{k}$ to obtain a shorter string $\bm{z}$ which is uniform and uncorrelated to any quantum adversary Eve. 
\end{enumerate} 

We provide a detailed template protocol for $N$-cycle NC inequalities in Figs.~\ref{fig:protocol_odd} and~\ref{fig:protocol_even}. We note that the step corresponding to measurement device characterization is not included in Figs.~\ref{fig:protocol_odd} and~\ref{fig:protocol_even} as it requires its own detailed explanation (see Sec.~\ref{sec:loopholefree}) and because of the SDI nature of the protocol, this step is already assumed to be performed prior to its execution. However, we note that this device characterization step effectively determines the parameter $\delta$ in Figs.~\ref{fig:protocol_odd} and~\ref{fig:protocol_even}.

Since we are interested in analyzing our protocol for a finite number of experimental rounds, we focus on empirical frequency distributions of outcomes instead of probabilities.
Here, the empirical frequency distribution $F_C(\mathcal{G})$ of an NC inequality over the graph $\mathcal{G}$ is defined as
\begin{equation}
    F_C(\mathcal{G}) := \frac{1}{n} \sum_{j = 1}^{n} \delta_{C_j,1},
\end{equation}
where $n$ is the number of rounds, the index $j$ corresponds to the $j$th round of the protocol, $C_j \in \lbrace \perp, 0, 1\rbrace$ $\forall j \in \lbrace 1, 2, \ldots, n\rbrace$ and $\delta_{C_j, 1}$  is the Kronecker delta function. Consequently, we can rewrite the violation of the NC inequality in terms of the empirical frequency distribution such that the violation lies within a confidence interval decided by a parameter $\delta \in (0,1)$ as $F_{C}(\mathcal{G}) \in \left[ q(\beta_Q - \delta),  q(\beta_Q + \delta)\right]$. In our protocol, we define $\delta=\delta_{\rm stat}+\delta_{\rm char}$, where $\delta_{\rm stat}$ accounts for finite statistical fluctuations and $\delta_{\rm char}$ quantifies deviations arising from imperfections in the measurement devices. In an actual experimental implementation the value of $\delta$ will be determined by the finite statistical fluctuations and the characterization of the measurement devices. We note that this formulation of $\delta$ in terms of statistical fluctuations and measurement imperfections is a modification of the notion that is typically employed in standard finite rate security proofs. We discuss this modification, especially $\delta_{\rm char}$ in detail in Sec.~\ref{sec:loopholefree}.

We can now also define the set of events in which the protocol does not abort by $\Omega = \lbrace \bm{C} | q \left(\beta_Q - \delta\right) \leq F_C(\mathcal{G}) \leq q \left(\beta_Q + \delta\right)\rbrace$, where $\bm{C} = C_1 C_2 \ldots C_n$.

The security of the random bit string produced by our protocol is guaranteed by the application of a strong quantum-proof randomness extractor, which essentially ensures that the resultant bit string of Alice and the initial random seed (required for the strong quantum-proof randomness extractor) are close to being uniformly distributed and both are also uncorrelated with an adversary holding any quantum side information. 

However, in order to apply the strong quantum-proof randomness extractor our protocol needs to produce enough randomness, which can be quantified by the conditional smooth min-entropy. To that end, we apply the entropy accumulation theorem (EAT), which provides a lower bound on the latter. Qualitatively, EAT ensures that the smooth min-entropy~\cite{RW04} produced by our protocol is at least
\begin{equation}
    H_{\min}^{\epsilon_s}(\textbf{A}|\textbf{X}, E) > nf - \mathcal{O}(\sqrt{n}), 
    \label{eq:eat_informal}
\end{equation}
where $\epsilon_s$ is the smoothing parameter, $\textbf{A}$ is the joint system of Alice over $n$ rounds, $\textbf{X}$ is the concatentation of all settings chosen by Alice in each round and $f$ is a function that depends on the conditional von-Neumann entropy evaluated for an observed value of $\beta_Q$. More details on conditional smooth min-entropy, EAT, the order parameter and how we evaluate the function $f$ can be found in Sec.~\ref{sec:methods}.

We also need to quantify the expected length of the input randomness, which can be done by noting that it is required for (i) selecting whether a round will be spot-check or generation round, and (ii) randomly selecting the measurement settings in the spot-check round. We note that since we employ strong quantum-proof randomness extractors, the length of the seed $\textbf{s}$ (which is private) should not be considered in the input randomness, as it can be utilized in the final random bit string $\bm{z}$. For our purposes, we employ the interval algorithm to generate the desired input randomness, such that its expected length is $l_{\rm in} = n \left[h(q) + q \log_2 N\right] + 6$~\cite{BRC20}, where $h(\cdot)$ is the binary Shannon entropy and the factor of $6$ appears because of the application of the interval algorithm. 
Consequently, we can define the net expansion rate as
\begin{equation}
    r' := \frac{H_{\min}^{\epsilon_s}(\textbf{A}|\textbf{X}, E) - l_{\rm in} - l_{\rm ext}}{n},
    \label{eq:rand_exp1}
\end{equation}
where $l_{\rm ext}$ is the amount of entropy loss incurred by the randomness extractor. There is a rich literature available on the construction and application of various strong quantum-proof randomness extractors~\cite{NA99, T01}. In particular, for universal hashing~\cite{RK05}, $l_{\rm ext}$ is given by
\begin{equation}
    l_{\rm ext} = 2\log_2\left(\frac{1}{\epsilon_{\rm ext}}\right),
\end{equation} 
where $\epsilon_{\rm ext}$ is the error parameter associated with randomness extraction. For $\epsilon_{\rm ext}=10^{-12}$, $l_{\rm ext}$ is approximately $79$ bits. Since $l_{\rm ext}$ is independent of the number of protocol rounds $n$, its contribution to the net expansion rate, $l_{\rm ext}/n$, becomes negligible for sufficiently large $n$. Therefore, for the remainder of this paper, we focus on the randomness expansion rate
\begin{equation}
    r := \frac{H_{\min}^{\epsilon_s}(\textbf{A}|\textbf{X}, E) - l_{\rm in}}{n}.
    \label{eq:rand_exp}
\end{equation}

\section{Numerical simulations}
\label{sec:results}

In this section we perform numerical simulations of our template protocols by considering $N = 5$ and $N = 4$ respectively. These two illustrative examples correspond to KCBS and CHSH inequalities, which are the two well known contextuality scenarios.

\subsection{Secure RE based on odd $N$-cycle NC inequalities}

\begin{figure}
    \centering
    \includegraphics[scale = 0.5]{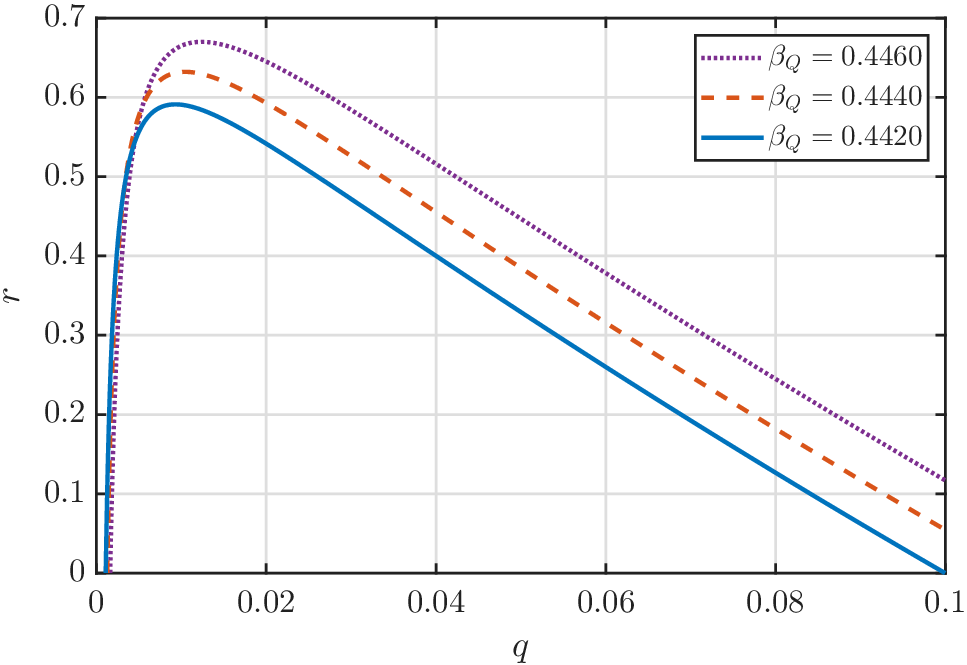}
    \caption{Expected net randomness expansion rate $r$ as a function of the spot-checking probability $q$ for $N = 5$. We choose $n = 10^{12}$ and $\delta = 3 \times 10^{-5}$ while the terms which appear in the order parameter in Eq.~\eqref{eq:eat_informal} are $\epsilon_{EA} = \epsilon_s = 10^{-6}$ and $\gamma = 1/\sqrt{n}$.}
    \label{fig:rate_vs_q}
\end{figure}

To illustrate a positive RE rate we first consider the odd-cycle NC scenario with $N=5$, corresponding to the KCBS inequality. 
We note that in this case the maximum NCHV bound is $\beta_{\rm NCHV} = 2/5$ and the maximum quantum value is $\beta_{\rm QT} = 1/\sqrt{5}$. 
In Fig.~\ref{fig:rate_vs_q} we plot $r$ as a function of the spot-checking probability $q$ for this case. We find that with the choice of parameters that we make in our protocol and optimizing over $q$, it is possible to obtain $r = 0.67$ bits per round when the violation is $\beta_Q = 0.4460$.

\begin{figure}
    \centering
    \includegraphics[scale = 0.5]{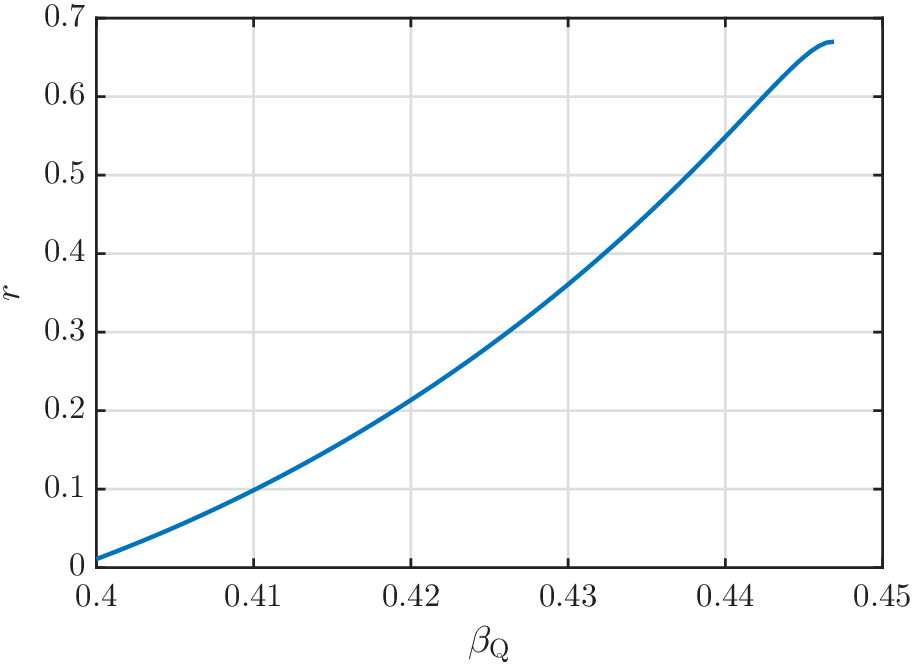}
    \caption{Expected net randomness expansion rate $r$ as a function of the violation of the KCBS inequality with $N = 5$. We choose $n = 10^{12}$, $q = 10^{-2}$ and $\delta = 3 \times 10^{-5}$ while the terms which appear in the order parameter in Eq.~\eqref{eq:eat_informal} are $\epsilon_{EA} = \epsilon_s = 10^{-6}$ and $\gamma = 1/\sqrt{n}$.}
    \label{fig:rate_vs_kcbs}
\end{figure}

Fixing the spot-checking probability $q = 10^{-2}$, we then study the dependence of RE rate on the observed violation of the KCBS inequality. The results are shown in Fig.~\ref{fig:rate_vs_kcbs}. As expected, the RE rate increases with the violation of the KCBS inequality and reaches a maximum value of approximately $0.67$ bits per round near the maximal quantum violation. This indicates that larger violations imply better security. Qualitatively, this can also be understood from the fact that the maximum violation of the NC inequality is obtained for a pure state, which further implies that the state of Alice is uncorrelated with an adversary.

\begin{figure}
    \centering
    \includegraphics[scale = 0.5]{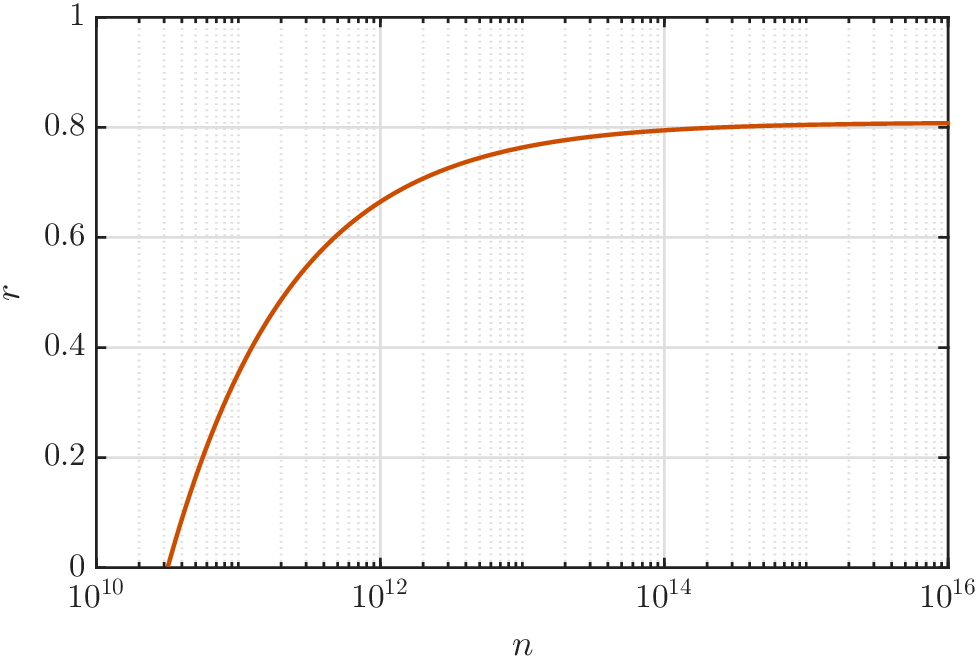}
    \caption{Expected net randomness expansion rate $r$ as a function of total number of rounds $n$ for $N=5$ cycle. We choose $q = 10^{-2}$, $\beta_Q = 0.4460$ and $\delta = 3 \times 10^{-5}$ while the terms which appear in the order parameter in Eq.~\eqref{eq:eat_informal} are $\epsilon_{EA} = \epsilon_s = 10^{-6}$ and $\gamma = 1/\sqrt{n}$.}
    \label{fig:rate_vs_n_kcbs}
\end{figure}

Next, we investigate the dependence of the RE rate on the total number of experimental rounds. In Fig.~\ref{fig:rate_vs_n_kcbs}, we plot $r$ as a function of $n$ for $q = 10^{-2}$ and $\beta_Q = 0.4460$. We observe that the rate increases with the number of rounds and approaches a saturation value of about $0.81$ bits per round for $n \gtrsim 10^{16}$.

\begin{figure}
    \centering
    \includegraphics[scale = 0.5]{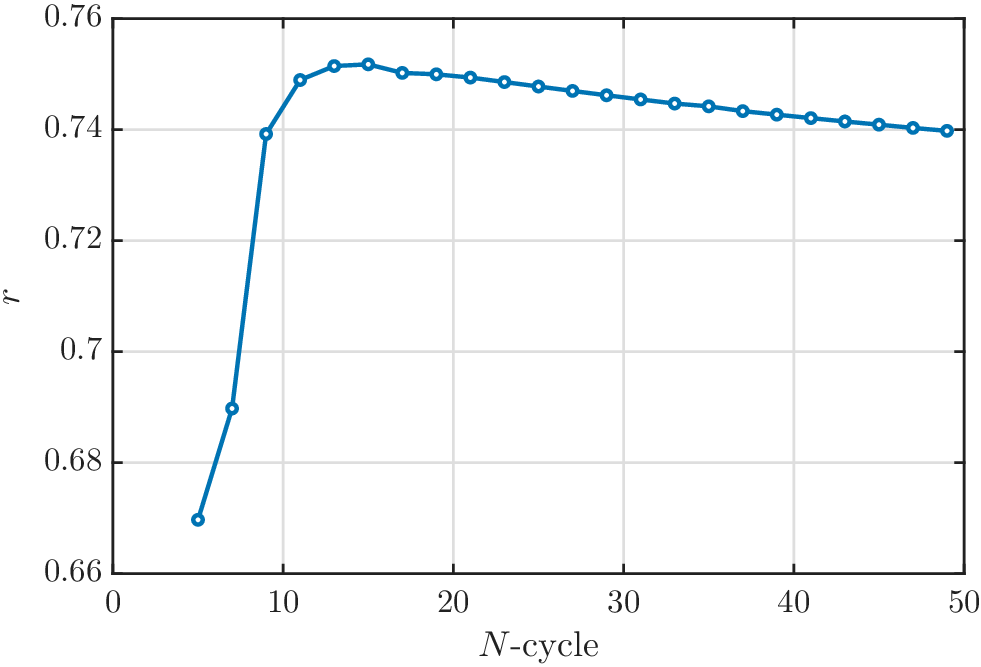}
    \caption{Expected net randomness expansion rate $r$ as a function of the number of odd-cycles. We choose $q = 10^{-2}$, $n=10^{12}$ and $\delta = 3 \times 10^{-5}$. For each odd $N$-cycle NC inequality, we take the maximum quantum violation $\beta_Q = \cos(\pi/N)/(1+\cos(\pi/N))$, while the terms which appear in the order parameter in Eq.~\eqref{eq:eat_informal} are $\epsilon_{EA} = \epsilon_s = 10^{-6}$ and $\gamma = 1/\sqrt{n}$.}
    \label{fig:rate_vs_N_kcbs}
\end{figure}

We also analyze the effect of increasing the number of cycles in the odd-cycle NC inequalities. In Fig.~\ref{fig:rate_vs_N_kcbs}, we plot the net randomness expansion rate as a function of $N$ for $n = 10^{12}$ and $q = 10^{-2}$. We find that the rate attains its maximum value of approximately $0.7517$ bits per round at $N=15$, after which $r$ decreases with increasing $N$. This behaviour can be understood as follows. For small values of $N$, the smooth min-entropy increases more rapidly than the input randomness cost $l_{\rm in}$, leading to an increase in the expansion rate. At $N=15$, the rate reaches its maximum. For larger values of $N$, the smooth min-entropy gradually approaches its upper bound of one bit per round, while the input randomness cost continues to increase with $q\log_2 N$. As a result, the RE rate decreases.

\subsection{Secure RE based on even $N$-cycle NC inequalities}

Next, we consider the even-cycle NC scenario, in particular $N = 4$. This scenario corresponds to the CHSH inequality, but instead of involving two non-communicating parties, it only involves a single party with non-disturbing measurements. Moreover, by considering the two outcome measurements (as in Eq.~\eqref{eq:meas_bell}), the maximum randomness that Alice can generate cannot be more than $1$ bit.

In this case the maximum NCHV is $\beta'_{\rm NCHV} = 3/4$ and the maximum quantum value is $\beta'_{\rm QT} = (2 + \sqrt{2})/4$. This is an interesting scenario in the sense that the CHSH inequality is known to play a pivotal role in DI-RE protocols, but its application in SDI-RE protocols remains largely unexplored. 

\begin{figure}
    \centering
    \includegraphics[scale = 0.5]{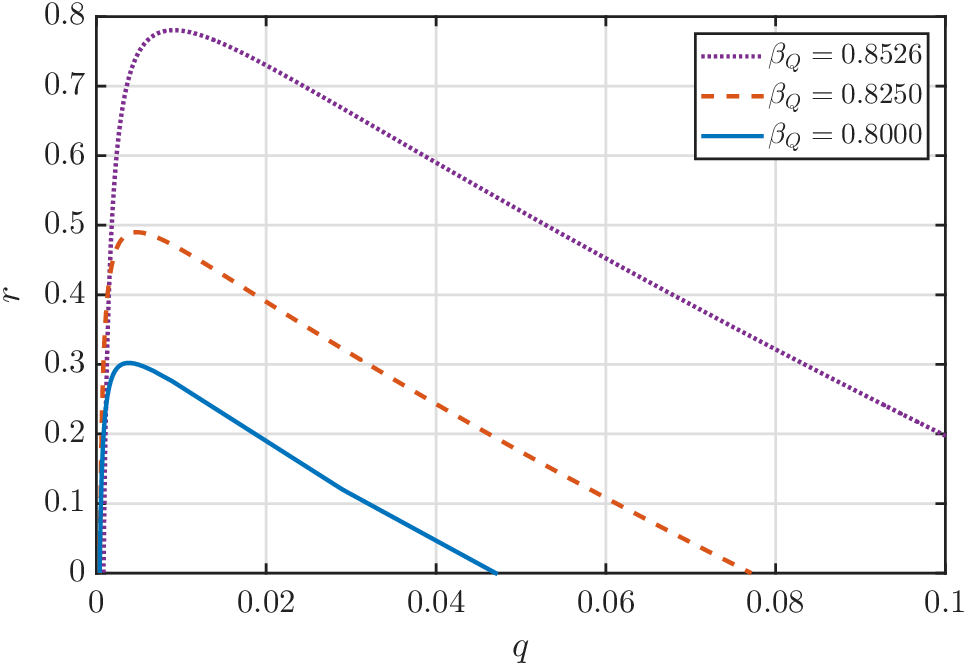}
    \caption{Expected net randomness expansion rate $r$ as a function of the spot-checking probability $q$ for $N = 4$. We choose $n = 10^{12}$, and $\delta = 3 \times 10^{-5}$ while the terms which appear in the order parameter in Eq.~\eqref{eq:eat_informal} are $\epsilon_{EA} = \epsilon_s = 10^{-6}$ and $\gamma = 1/\sqrt{n}$.}
    \label{fig:rate_vs_q_bell}
\end{figure}

\begin{figure}
    \centering
    \includegraphics[scale = 0.5]{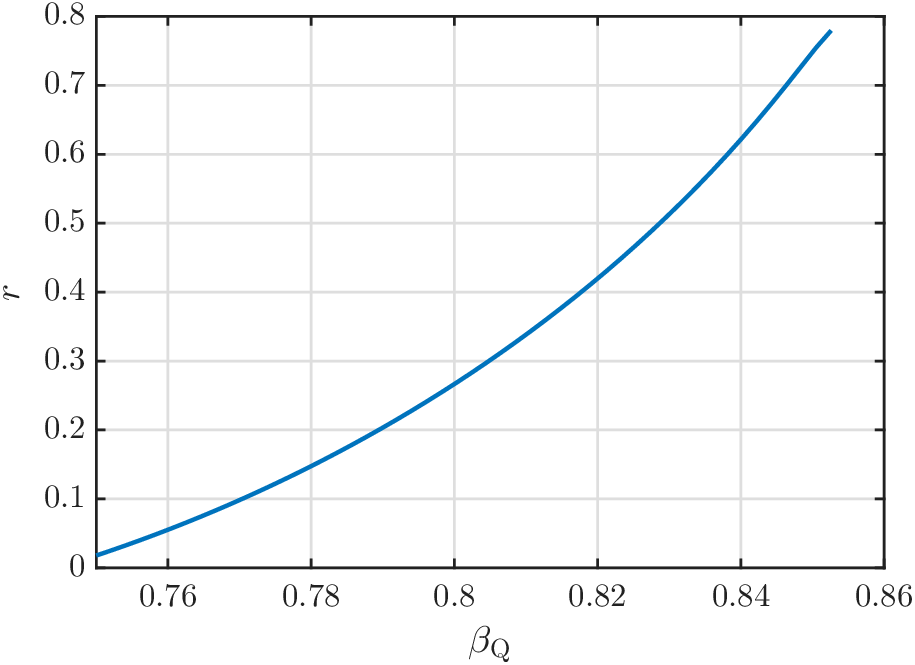}
    \caption{Expected net randomness expansion rate $r$ as a function of the violation of the NC inequality with $N = 4$. We choose $n = 10^{12}$, $q = 10^{-2}$ and $\delta = 3 \times 10^{-5}$ while the terms which appear in the order parameter in Eq.~\eqref{eq:eat_informal} are $\epsilon_{EA} = \epsilon_s = 10^{-6}$ and $\gamma = 1/\sqrt{n}$.}
    \label{fig:rate_vs_bell}
\end{figure}

In Fig.~\ref{fig:rate_vs_q_bell}, we plot $r$ as a function of the spot-checking probability $q$. We find that with an appropriate choice of parameters, it is possible to obtain $r = 0.78$ bits per round when the observed violation is $\beta_Q = 0.8526$. Similar to the $N=5$ case, we fix $q = 10^{-2}$ and study the dependence of the randomness expansion rate on the observed violation. The results are shown in Fig.~\ref{fig:rate_vs_bell}. 

\begin{figure}
    \centering
    \includegraphics[scale = 0.5]{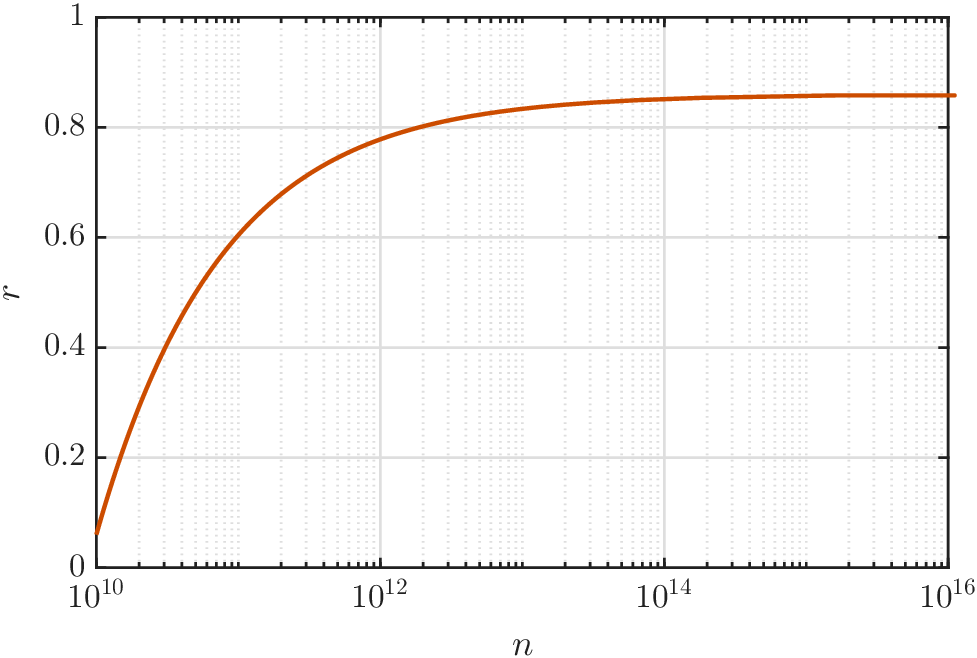}
    \caption{Expected net randomness expansion rate $r$ as a function of total number of rounds $n$ for $N=4$ cycle. We choose $q = 10^{-2}$, $\beta_Q = 0.8526$ and $\delta = 3 \times 10^{-5}$ while the terms which appear in the order parameter in Eq.~\eqref{eq:eat_informal} are $\epsilon_{EA} = \epsilon_s = 10^{-6}$ and $\gamma = 1/\sqrt{n}$.}
    \label{fig:rate_vs_n_chsh}
\end{figure}

Next, we investigate the dependence of the net randomness expansion rate on the total number of experimental rounds. In Fig.~\ref{fig:rate_vs_n_chsh}, we plot $r$ as a function of $n$ for $q = 10^{-2}$ and $\beta_Q = 0.8526$. We observe that the rate increases with the number of rounds and gradually approaches a saturation value of about $0.86$ bits per round for $n \gtrsim 10^{16}$. 

\begin{figure}
    \centering
    \includegraphics[scale = 0.5]{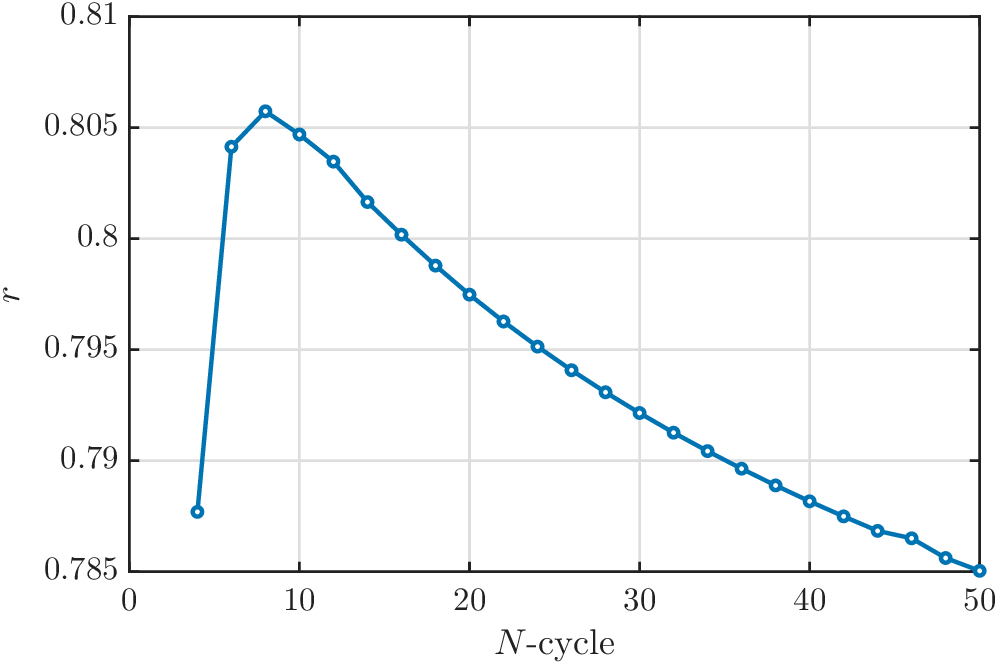}
    \caption{Expected net randomness expansion rate $r$ as a function of the number of even-cycles. We choose $q = 10^{-2}$, $n=10^{12}$ and $\delta = 3 \times 10^{-5}$. For each even-cycle NC inequality, we take the maximum quantum violation $\beta_Q = \left(1 + \cos \left(\pi/N\right)\right)/2$, while the terms which appear in the order parameter in Eq.~\eqref{eq:eat_informal} are $\epsilon_{EA} = \epsilon_s = 10^{-6}$ and $\gamma = 1/\sqrt{n}$.}
    \label{fig:rate_vs_N_chsh}
\end{figure}

Finally, we investigate the effect of increasing the number of cycles in the even-cycle NC inequalities. As shown in Fig.~\ref{fig:rate_vs_N_chsh}, for $n=10^{12}$ and $q=10^{-2}$, the randomness expansion rate reaches a maximum of approximately $0.8057$ bits per round at $N=8$. Beyond this point, the increase in the input randomness cost $l_{\rm in}$, which grows with $q\log_2 N$, exceeds the gain in the smooth min-entropy, causing the net RE rate to decrease.

\section{Detailed security analysis}
\label{sec:methods}

In this section we provide a more detailed security analysis of our protocol for finite statistics. 

Let us first define $\textbf{A} = A_1 A_2 \ldots A_n$ as the joint system of Alice over $n$ experimental rounds, where, in each round $j$ her system is denoted by $A_j$ and the concatenation of $j$ number of rounds is denoted by $\textbf{A}_{j}$. In a similar fashion, we define $\textbf{X} = X_1 X_2 \ldots X_n$ which is the concatenation of the settings chosen by Alice such that $X_j$ denotes the setting chosen in the $j$th round and the concatenation of $j$ number of rounds is denoted by $\textbf{X}_{j}$. The same convention applies to the register $T$.
Furthermore, let $\rho_{\bm{z}\bm{s}E}$ denote the quantum state describing the final output of the protocol, where $E$ is the system of Eve corresponding to any side information that may be available to her. 

We can now define a strong quantum-proof randomness extractor as follows.

\begin{definition}
    (Strong quantum-proof randomness extractors)
    \label{def:qpre}
    A function $R_{\rm ext}: \lbrace 0,1 \rbrace^{|\bm{k}|} \times \lbrace 0,1\rbrace^{|\bm{s}|} \rightarrow \lbrace 0,1\rbrace^{|\bm{z}|}$ is a  quantum-proof $(k, \epsilon_{\rm ext} + 2\epsilon_s)$-strong randomness extractor if for all cq-states $\rho_{\textbf{A}E}$ with smooth min-entropy $H_{\min}^{\epsilon_s}(\textbf{A}|\textbf{X},E) \geq k$ (for more information see Ref.~\cite{RW04}) and for some $\epsilon_s > 0$ it satisfies
    \begin{equation}
        \frac{1}{2}\norm{\rho_{\bm{z}\bm{s}E} - 2^{-(l + |\bm{s}|)} \mathds{1}_l \otimes \mathds{1}_{|\bm{s}|} \otimes \rho_E} \leq \epsilon_{\rm ext} + 2 \epsilon_s,
    \end{equation}
    where $\epsilon_{\rm ext}$, $\epsilon_s \in (0,1)$, $\mathds{1}_l$ denotes an identity matrix of dimension $l \times l$, $\rho_E$ is a reduced state of Eve which is uncorrelated with Alice and the seed and $\norm{B} = {\rm tr}(\sqrt{B^\dagger B})$ denotes the trace norm (or trace distance).
\end{definition}

The application of a strong quantum-proof randomness extractor to the bit string $\bm{k}$ then guarantees that the resultant bit string $\bm{z}$ is close (in trace distance) to a uniform distribution which is uncorrelated to any adversary's system. Consequently, our protocol satisfies a strong notion of security which allows the resultant random bit string to be composed with any other cryptographic protocol without compromising its security. Such a notion of security is termed as universally composable security and is formally defined as follows. 

\begin{definition}
    (Universally composable security)
    \label{def:sec}
    A randomness expansion protocol is $(\epsilon_{\rm sound}, \epsilon_{\rm comp})-$secure if 
    \begin{equation}
        \frac{1}{2} {\rm Pr}\left[\Omega\right] \norm{\rho_{\bm{z}\bm{s}E} - 2^{-(l + |\bm{s}|)}\mathds{1}_l \otimes \mathds{1}_{|\bm{s}|} \otimes \rho_E} \leq \epsilon_{\rm sound},
        \label{eq:sec_def1}
    \end{equation}
    where ${\rm Pr}\left[\Omega\right] \geq 1 - \epsilon_{\rm comp}$ is the probability for the protocol to not abort and $\epsilon_{\rm comp}$, $\epsilon_{\rm sound} \in (0,1)$ denote the completeness and soundness of the protocol respectively.
\end{definition}

For a universally composable secure RE, our scheme should satisfy the condition given in Eq.~\eqref{eq:sec_def1}. More specifically, i) our protocol should have a small probability to abort such that $\epsilon_{\rm comp} \ll 1$ and ii) the resultant output string $\bm{z}$ and the seed $\bm{s}$ should be close to being uniform and uncorrelated to any adversary's (quantum) system $E$, such that $\epsilon_{\rm sound} \ll 1$. The aforementioned two conditions are termed as completeness and soundness respectively.

Next, we show how the completeness parameter $\epsilon_{comp}$ can be evaluated based on other parameters used in the protocol. 

Following Theorem $1$ of Ref.~\cite{LLR21} (and likewise Theorem $2$ of Ref.~\cite{WPN23}) we can choose the completeness parameter of the protocol as
\begin{equation}
    \epsilon_{\rm comp} = 1 - F(n, q(1 - \beta_Q), \lfloor nq\left[1 - (\beta_Q - \delta) \right]\rfloor),
    \label{eq:comp}
\end{equation}
where $F(n,p,k) = \Phi\left({\rm sign}\left(\frac{k}{n} - p\right)\sqrt{2nD\left(\frac{k}{n},p\right)}\right)$, such that $\Phi(x)$ is the error function defined as $\Phi(x) = \frac{1}{\sqrt{2 \pi}} \int_{-\infty}^{x} e^{-t^2/2} dt$, $D(x,y) = x \ln (\frac{x}{y}) + (1 - x)\ln(\frac{1 - x}{1 - y})$ and $\lfloor \cdot \rfloor$ denotes the floor function. This choice allows us to satisfy the completeness condition for universally composable security.

As an example, consider the choice $\delta = 3 \times 10^{-5}$, $N = 5$, $\beta_Q = 0.4460$, $n = 10^{12}$ and $q = 1.22 \times 10^{-2}$. With this we already achieve $\epsilon_{\rm comp} \approx  10^{-6}$.

Next, we elucidate how to bound the soundness parameter of our protocol. 

Firstly, in order to apply the extractor we need to certify that our protocol produces enough randomness which is quantified by the conditional smooth min-entropy. 

Evaluating the conditional smooth min-entropy is generally not an easy task, but due to recent efforts, particularly in the form of entropy accumulation theorem (EAT), it can be lower bounded in terms of the conditional von-Neumann entropy of the worst single round of the protocol. Before stating the EAT, recall that it applies to any sequence of rounds satisfying two structural conditions~\cite{DF19, DFR20}: the score $C_j$ must be a deterministic function of $A_j$, $T_j$ and $X_j$, and the Markov chain condition $I(\textbf{A}_{j-1} : T_jX_j \mid \textbf{T}_{j-1}\textbf{X}_{j-1} E) = 0$ $\forall j \in \lbrace 1, \ldots, n \rbrace$
must hold. In our protocol, $C_j$ is assigned directly from $A_j$, $T_j$, and $X_j$ by the case distinction in Figs.~\ref{fig:protocol_odd} and~\ref{fig:protocol_even}, and $T_j$ and $X_j$ are sampled from Alice's private random seed independently of every previous round. Furthermore, they are sampled before the measurements of round $j$ are performed. As a result $T_j$ and $X_j$ cannot depend on $\textbf{A}_{j-1}$, $\textbf{T}_{j-1}$, $\textbf{X}_{j-1}$, or on $E$. As a result, the conditions for applying EAT are satisfied in our protocol.

\begin{definition}
    (Entropy accumulation theorem)
    \label{def:eat}
    Let $\Omega$ denote the event in which the protocol for QRE does not abort such that the final state at the end of the protocol is $\rho_{|\Omega}$. Let $f$ be an affine function, termed as a min-tradeoff function, that lower bounds the single round von-Neumann entropy for any quantum realization that achieves $\beta_Q > \beta_{\rm NCHV}$, then for $\epsilon_s$, $\epsilon_{EA}$, $\gamma \in \left(0,1\right)$, our QRE protocol accepts with probability $\epsilon_{EA}$ 
    \begin{equation}
        \begin{aligned}
            H_{\min}^{\epsilon_s}(\textbf{A}|\textbf{X}, E)_{\rho_{|\Omega}} > nf - &\frac{1}{\gamma} \left[1 - 2 \log_2(\epsilon_{EA} \epsilon_{s})\right]\\
            &- n\left[\gamma V + \gamma^2 K\right], 
        \end{aligned}
        \label{eq:EAT}
    \end{equation}
    where $H_{\min}^{\epsilon_s}(\textbf{A}|\textbf{X}, E)_{\rho_{|\Omega}}$ is the conditional smooth min-entropy with smoothing parameter $\epsilon_s$ over the joint systems of Alice $\textbf{A}$ conditioned on her joint settings $\textbf{X}$ and the quantum side information $E$ held by an adversary. Here, the correction terms $V$ and $K$ are defined as
    \begin{equation}
        \begin{aligned}
            V &:= \frac{\ln 2}{2} \left(\log_2 9 + \sqrt{{\rm var}(f) + 2}\right)^2,\\
            K &:= \frac{1}{6(1 - \gamma)^3 \ln 2} 2^{\gamma \left(1 + \max(f) - \min(f)\right)}\\
                &\quad \times \ln (2^{\left(1 + \max(f) - \min(f)\right)} + e^2),
        \end{aligned}
    \end{equation}
    where $\max(\cdot)$, $\min(\cdot)$ and ${\rm var}(\cdot)$ denote the maximum, minimum and variance of the function respectively.
    
\end{definition}

The EAT guarantees that either our protocol aborts with a probability at most $1-\epsilon_{EA}$ or generates a certain amount of smooth min-entropy. In the case when the protocol does not abort, following Definition~\ref{def:qpre} we can choose $\epsilon_{\rm sound} = \max \lbrace \epsilon_{EA}, \epsilon_{\rm ext} + 2\epsilon_s\rbrace$. For our purposes we choose $\epsilon_{\rm sound} = \epsilon_{EA} = \epsilon_{\rm ext} + 2\epsilon_s = 10^{-6}$.

We note that an even stronger version of EAT, termed as generalized EAT, has been derived using a more general model in which the adversary can update their quantum side information in each round. In our protocol there is only a single party Alice who does not reveal any side information and as such we do not need to consider the generalized version of EAT which would offer worse rates.

Finally, in order to exhibit randomness expansion, the QRE protocol should output a random bit string having length $l = \abs{\bm{z}}$ greater than the length of the random bit string that is consumed for the operation of the protocol, which we denote by $l_{\rm in}$.

In order to apply EAT a reliable method is required to evaluate von-Neuamnn entropies conditioned on any quantum side information available to any adversary for arbitrary violations of the NC inequality. 
To that end, we describe a method to compute the conditional von-Neumann entropy using the semi-definite programming approach developed in Ref.~\cite{BFF21-2}. Although this technique holds for scenarios in which both the measurement and preparation devices are uncharacterized (or untrusted), a modification of the same was also recently developed in Ref.~\cite{AHN23} which allows for the measurement devices to be well characterized and trusted while the source is not. Therefore, to compute the worst single round conditional von Neumann entropy we employ the latter SDP approach which is well suited to our SDI scenario. In this approach, the conditional von Neumann entropy can be evaluated as
\begin{widetext}
\begin{equation}
\begin{aligned}
    H(A|X = x^*, E) =  c_m + &\underset{\sigma, \lbrace \zeta_i^{a}, \eta_i^a, \theta_i^a\rbrace_{a, i}}{\min} \sum_{i = 1}^{m} \sum_{a = 0}^{1} \frac{w_i}{t_i \ln 2} \text{tr}\left[\left(M_{a|x^*} \otimes \mathds{1}_E \right) \left(\zeta_i^a + \zeta_i^{a\dagger} + \left(1 - t_i\right)\eta_i^a\right) + t_i \theta_i^a\right]\\
    \qquad \text{s.t.}\quad &\text{tr}\left(\sigma\right) = 1,\\ 
    &\beta = \beta_{Q} - \delta,\\
    &\Gamma_{a,i}^1 := \begin{pmatrix}
    \sigma & \zeta_i^a \\
    \zeta_i^{a \dagger} & \eta_i^a
    \end{pmatrix} \geq 0, ~ \Gamma_{a,i}^2 := \begin{pmatrix}
        \sigma & \zeta_i^{a \dagger} \\
        \zeta_i^a & \theta_i^a
    \end{pmatrix} \geq 0 ~ \forall a, i,
\end{aligned}
\label{eq:sdp_primal}
\end{equation}
\end{widetext}
where $m \in \mathbb{N}$, $t_1, \ldots, t_m$ and $w_1, \ldots, w_m$ are the nodes and weights of a $m$ point Gauss-Radau quadrature on $[0, 1]$ with an endpoint $t_m = 1$ and $c_m = \sum_{i = 1}^{m}\frac{w_i}{t_i \ln 2}$. Here, the second constraint in Eq.~\eqref{eq:sdp_primal} denotes the worst-case scenario when the contextual correlations are observed closer to the non-contextual value $\alpha(\mathcal{G})$. This is taken to evaluate the worst-case RE rate.

The aforementioned SDP is also particularly useful in constructing the min-tradeoff function $f$, which is a necessary ingredient in Definition.~\ref{def:eat}.

In order to construct the min-tradeoff function we consider the dual solution of the SDP in Eq.~\eqref{eq:sdp_primal}, which lower bounds the von-Neumann entropy as
\begin{equation}
    H(A|X = x^*, E) \geq c_m - \bm{\lambda}_Q\cdot \bm{\omega}_Q,
\end{equation}
where $\bm{\omega}_Q = (1, \beta_Q)$ are the constraint values and $\bm{\lambda}_Q$ are the dual solutions to the SDP. From the dual solution, we can construct an affine function $g$ defined as
\begin{equation}
    g(\bm{e}_c) = (1 - q)(c_m - \bm{\lambda}_Q\cdot \bm{e}_c),
\end{equation}
where $\bm{e}_c$ is the probability distribution which has the $c$th entry as one while all the other entries are zero. The properties of the function $g$ can be related to the required properties of the min-tradeoff function $f$ as follows:
\begin{equation}
    \begin{aligned}
        \max(f) &= \max(g),\\
        \min(f) &\geq \min(g),\\ 
        {\rm var}(f) &\leq \frac{\left(\max(g) - \min(g)\right)^2}{q},
    \end{aligned}  
\end{equation}
where $\max(g) = (1 - q)(c_m - \lambda_{\min})$ and $\min(g) = (1 - q)(c_m - \lambda_{\max}))$ and we have used the shorthand notation $\lambda_{\min} = \min(\bm{\lambda}_Q)$ and $\lambda_{\max} = \max(\bm{\lambda}_Q)$. Following Ref.~\cite{BRC20} the min-tradeoff function can be written as
\begin{equation}
    f(\bm{\omega}_Q) = (1 - q) (c_m - \bm{\lambda}_Q \cdot \bm{\omega}_Q).
\end{equation}

Following the construction of the min-tradeoff function, we can employ EAT to obtain the net randomness expansion rate. In Tables~\ref{tab:param1} and~\ref{tab:param2} we list the parameters that can be used to illustrate RE from our scheme secure against quantum adversaries.

\setlength{\tabcolsep}{7.5pt} 

\begin{table}
    \caption{Parameters for observing the expected randomness expansion for $N=5$. We take the total experimental rounds $n=10^{12}$ and set $\epsilon_{\rm EA}=10^{-6}$.}
    \centering
    \begin{tabular}{ccccc}
        \hline
        \textbf{$\delta$} & \textbf{$\epsilon_{\rm comp}$} &  \textbf{$q$} & \textbf{$\beta_Q$} & \textbf{$r$} \\
        \hline
        \hline
        $3 \times 10^{-5}$ & $10^{-5}$ & $0.91\times 10^{-2}$ & $0.4420$ & $0.5911$ \\
        $3 \times 10^{-5}$ & $10^{-6}$  & $1.22 \times 10^{-2}$ & $0.4460$ & $0.6699$ \\
        \hline
    \end{tabular} 
     \label{tab:param1}
\end{table}

\begin{table}
    \caption{Parameters for observing the expected randomness expansion for $N=4$. We take the total experimental rounds $n=10^{12}$ and set $\epsilon_{\rm EA}=10^{-6}$.}
    \centering
    \begin{tabular}{ccccc}
        \hline
        \textbf{$\delta$} & \textbf{$\epsilon_{\rm comp}$} & \textbf{$q$} & \textbf{$\beta_Q$} & \textbf{$r$} \\
        \hline
        \hline
        $3\times 10^{-5}$ & $10^{-9}$ & $0.63 \times 10^{-2}$ & $0.8400$ & $0.6368$ \\
        $3\times 10^{-5}$ & $10^{-14}$  & $0.89\times 10^{-2}$ & $0.8526$ & $0.7802$ \\
        \hline
    \end{tabular} 
     \label{tab:param2}
\end{table}

\section{Characterization of measurement devices}
\label{sec:loopholefree}

In this section we detail the necessary characterization of the measurement device that needs to be performed prior to executing the protocol in Figs.~\ref{fig:protocol_odd} and~\ref{fig:protocol_even}. In particular we note that it is not necessary that the projectors be exactly equal to the ones mentioned in Eq.~\eqref{eq:projectors}. In fact, any other choice of projectors (such that their dimension is known) also works provided the conditions listed below are satisfied. It is in this sense that we term that our protocol requires only a partial characterization of the measurement devices. For more details on these conditions and how they can be experimentally verified we refer to Ref.~\cite{HXA23}.

These conditions include testing 
\begin{enumerate}
    \item \textbf{Repeatability}: The measurements involved in the NC inequality should be projective such that $\Pi_i^2 = \Pi_i$ for all $i$. At the level of observed statistics this condition translates to $\langle \Pi_i^2\rangle = \langle \Pi_i \rangle$ for all $i$. This condition can be ensured by evaluating the following quantity for a fixed state,
    \begin{equation}
        R_i := \sum_a p(a,a|i,i),
        \label{eq:repeatability}
    \end{equation}
     where $p(a,a|i,i)$ is the probability of observing the outcome $a$ twice when the same measurement corresponding to setting $x = i$ is performed sequentially twice. For a finite statistical analysis, the probabilities will be replaced by the corresponding frequency distribution. 

     Ideally, for projective measurements, $R_i = 1$, but due to even small imperfections or presence of noise in the devices, achieving this value might be impossible. As such, we consider the deviation from ideal case by defining 
     \begin{equation}
         \epsilon_{\rm rep} := \frac{1}{N}\sum_{i = 1}^{N}(1 - R_i^{\rm obs}),
     \end{equation}
     where $R_i^{\rm obs}$ is the experimentally observed value of the quantity $R_i$. The quantity $\epsilon_{\rm rep}$ can then be understood as the average deviation of the repeatability condition from the ideal statistics. To ensure higher security we require $\epsilon_{\rm rep} \ll 1$. However, the exact upper bound on $\epsilon_{\rm rep}$ will depend on the quality of the experiment which in turn will also determine the trust level or the security of the protocol. 

     \item \textbf{Orthogonality}: The measurements involved in the NC inequality should also ensure compatibility following the underlying graph $\mathcal{G}$. Since the measurements are assumed to be projective, this condition translates to testing the orthogonality of the projectors $\lbrace \Pi_i \rbrace$ involved in the NC inequality. This can be achieved for evaluating the following quantities for a fixed state,
     \begin{equation}
         O_{i,j} := \sum_{a, b = 0}^{1} |p(a,b|i,j) - p(b,a|j,i)|,
     \end{equation}
     and 
     \begin{equation}
         O'_{i,j}:= p(1,1|i,j),
     \end{equation}
     where the quantity $O_{i,j}$ captures the fact that ideally for orthogonal projective measurements changing the sequence of measurements should not affect the statistics, while the quantity $O'_{i,j}$ captures the fact that ideally two orthogonal projective measurements cannot yield the outcome $1$ when measured sequentially. Ideally for perfect orthogonality both the quantities $O_{i,j}$ and $O'_{i,j}$ should be zero.

     As before for a finite statistical analysis, the probabilities will be replaced by the corresponding frequency distributions and we define the average deviation from the ideal case by defining 
     \begin{equation}
        \begin{aligned}
            \epsilon_{\rm orth} &:= \frac{1}{N}\sum_{i = 1}^{N} O^{\rm obs}_{i, i+1},\\
            \epsilon'_{\rm orth} &:= \frac{1}{N} \sum_{i = 1}^{N} O'^{\rm obs}_{i, i+1},
        \end{aligned}
     \end{equation}
     where $O^{\rm obs}_{i, i+1}$ and $O'^{\rm obs}_{i, i+1}$ are the experimentally observed values of the quantities $O_{i, i+1}$ and $O'_{i, i+1}$ respectively, and the summation in the subscripts is taken modulo $N$.
\end{enumerate}

Next, we quantify the deviation from the ideal measurement conditions required for a violation of the NC inequality by defining
\begin{equation}
    \delta_{\rm char}:= \frac{1}{3}(\epsilon_{\rm rep} + \epsilon_{\rm orth} + \epsilon'_{\rm orth}).
\end{equation}

We can then define the total deviation parameter as
\begin{equation}
    \delta := \delta_{\rm stat} + \delta_{\rm char}
\end{equation}
where $\delta_{\rm stat}$ accounts for finite statistical fluctuations. The total deviation $\delta$ determines the confidence interval for the empirical frequency distribution in our RE protocol. Consequently, it also affects the completeness condition (see Defintion~\ref{def:sec} and Eq.~\eqref{eq:comp}), which determines the probability that the protocol aborts.

In the significantly loophole-free experiment in Ref.~\cite{HXA23}, the authors report the quantity $\epsilon_{\rm orth} \approx 10^{-2}$ (note that this quantity is denoted as $\delta_{ij}$ in their results), which is an order of magnitude larger than $\epsilon_{\rm rep}$, $\epsilon'_{\rm orth}$ as reported in their results. As a consequence, it is apparent that $\epsilon_{\rm orth}$ would constitute a major limiting factor to implementing our scheme. For the purposes of illustrating our scheme, we instead take both $\delta_{\rm char}$ and $\delta_{\rm stat}$ in the order of $10^{-5}$, which we note is significantly smaller than $\epsilon_{\rm orth}$. We believe this value will be achievable in future experiments.

\section{Conclusions}
\label{sec:conc}

We have presented a template SDI-RE protocol based on NC inequalities that requires only a partial characterization of the measurement device, while the preparation source is completely uncharacterized. Furthermore, our protocol requires only a single localized party for RE as opposed to several other schemes that necessitate at least two non-communicating parties. We have also shown that our protocol is secure against computationally unbounded quantum adversaries under the effects of finite statistics using the definition of universal composability. 

The NC inequalities we consider as constraints in our template protocol also subsume Bell inequalities under the condition that the measurement devices are partially characterized and the quantum system is in possession with a single localized party. Such scenarios, from the perspective of quantum cryptography, have been largely unexplored.
Our results highlight that such scenarios, which are experimentally more accessible than Bell non-local scenarios, can still offer secure RE rates under finite statistics.

For the finite rate security analysis we leverage the recently developed SDP technique of Ref.~\cite{AHN23} and the EAT~\cite{DFR20}. 
This highlights the versatility of the SDP approach introduced in Ref.~\cite{AHN23}, demonstrating its applicability to a range of cryptographic scenarios with varying security assumptions. While this answers in affirmative one of the open questions in Ref.\cite{AHN23}, it also opens up several new research avenues. Particularly, in the context of alternative SDI-RE scenarios where it is not always straightforward on how to apply EAT. While our template protocol utilized the violation of NC inequalities as a constraint, this can be generalized to consider constraints like energy constraints~\cite{HWCGP17}, Leggett-Garg inequalities~\cite{nath2024single} or quantum interference effects.

The downside of our scheme is that it functions under the assumption of memoryless devices. This assumption is necessary for a loophole-free test of any NC inequality and is a drawback in all cryptographic schemes based on tests of quantum contextuality. Recent results indicate that it may be possible to remove this assumption completely (based on Ref.~\cite{ABC24}), but at the expense of more resources. As such our results also motivate future research into removing or mitigating the issue of memoryless devices in loophole-free tests of NC inequalities. 

We note that some earlier works have also analyzed the task of RE using quantum contextuality, but either the security is only proven against classical adversaries for an asymptotically infinite number of experimental rounds~\cite{UZZ13} or the underlying security formalism used is known to be sub-optimal~\cite{MS17} and requires perfect compatibility among certain measurements~\cite{UZZ20}, which in general cannot be ensured in experimental tests of quantum contextuality. To combat the latter issue, the authors of Ref.~\cite{UZZ20} proposed a modified version of the KCBS inequality in which the quantum violation is penalized by an appropriate amount if the required compatibility relations are not perfectly satisfied in the experiment. While these modifications may be well motivated for experimental tests of quantum contextuality, they may not be fully justified in the context of security of RE. In particular the security formalism used~\cite{UZZ20} necessitates perfect compatibility but this cannot be guaranteed by changing the quantum violation alone. In a recent work~\cite{LR25}, the authors develop novel NC inequalities which incorporates imperfect compatibility among the observables and certify maximal randomness. However, these NC inequalities require sequential (or simultaneous) measurements of more than two observables which increases the complexity of an experimental realization.

On the other hand, our template protocol does not require any such modifications of the NC inequalities or development of new NC inequalities. Instead, any deviation of experimental statistics from a perfectly loophole-free test of quantum contextuality, which includes certifying repeatability and compatibility, can be incorporated in the probability to abort the protocol. This is in contrast with earlier schemes. Additionally, our scheme follows the stronger security definition of universal composability~\cite{C01} as opposed to some earlier works.

\section{Code Availability}
All the code and data necessary to reproduce the results of this paper are available in Ref.~\cite{G26}.

\section{Acknowledgment}
D.G. and N.J acknowledge financial support from the IIT Mandi HTRA fellowship. J.S. acknowledges financial support from the IIT Mandi seed grant project No. IITM/SG/JSN/168. J.S. also acknowledges funding support from the National Quantum Mission (NQM), an initiative of the Department of Science and Technology (DST), Govt. of India with project number IITM/DST-NQM/JSN/615. J.S. also acknowledges the support provided by the Foundation for QC Innovation (FQCI), DST-NQM T-Hub at IISc Bengaluru, in facilitating this project.


%

\end{document}